\documentclass[twocolumn,showpacs,aps,prl,a4paper]{revtex4}
\usepackage{amsmath,graphicx}

\begin{document}

\title{Jaynes-Cummings dynamics with a matter wave oscillator}
\author{Klaus M\o lmer}
\affiliation{ QUANTOP, Danish National Research Foundation Center for
Quantum Optics,
\\Department of Physics and Astronomy,
University of Aarhus \\
DK-8000 \AA rhus C, Denmark}

\begin{abstract}
We propose to subject two Bose-Einstein condensates to a periodic potential,
so that one condensate undergoes the Mott insulator transition to a state
with precisely one atom per lattice site. We show that 
photoassociation of heteronuclear molecules within each lattice site is 
described by
the quantum optical Jaynes-Cummings Hamiltonian. In analogy with
studies of this Hamiltonian with cavity fields and trapped ions, 
we are thus able to engineer quantum optical states of 
atomic matter wave fields and we are able to reconstruct these states 
by quantum state tomography.
\end{abstract}


\maketitle

The Jaynes-Cummings model was introduced to describe the 
resonant interaction between
an atom and a single mode of the  quantized radiation field
\cite{Cummings}, and with
incorporation of pumping and relaxation mechanisms, it constitutes
the corner stone for the quantum treatment of maser and laser action
\cite{ScullyLambHaken}. The unitary Jaynes-Cummings dynamics has 
been succesfully demonstrated in the micro maser \cite{Micromaser}, 
in which atoms in Rydberg excited 
states are injected into a high-Q cavity with a single mode which is 
resonant with an atomic transition.
This system has been used to demonstrate fundamental 
aspects of quantum theory such as the oscillatory exchange of energy 
between the atom and the field, sub-poissonian photon statistics, 
entanglement, and decoherence \cite{QED,ENS-cat}.
Another implementation of the Jaynes-Cummings Hamiltonian is in the
description of the quantized motion of an ion in a harmonic trap
\cite{Ions}, for which a classical laser field 
drives transitions in which the ion is excited and
simultaneously one quantum of energy is extracted from the motional state.
The ion system is characterized by very long lifetimes of the oscillator
state, and hence the Jaynes-Cummings dynamics has been studied in great
detail, and numerous examples of state synthesis and detection have been
presented \cite{Meekhof,Leibfried}.

In a quantum optical perspective, what characterizes a Bose-Einstein
condensate is its single-mode character and coherence properties
\cite{Atomcoh}. 
The bosonic commutator relation applies for atoms as well as for
photons and phonons, and if one neglects interactions, 
many features of single mode
fields are thus common for atoms, light and harmonic motion. 
This has of course led to
various proposals for generation of atomic states with properties 
similar to the ones of light, e.g., phase and number squeezed 
states and Schr\"odinger cat states. 
By clever use of the atomic interactions as an adjustable
non-linear element, strong quantum effects  are foreseen
\cite{Anders}, and in combination with the long storage 
times for atoms, these effects are potentially much more useful,
for atoms than for light, and indispensable, e.g., for the 
improvement of atomic clocks \cite{Clocks}.

In this Letter we present a quantum optical analysis of
the dynamics of a pair of atomic condensates which have been
trapped in a periodic potential, so that one species has experienced
the Mott insulator transition to a state with exactly one atom
per lattice site \cite{Jaksch98,Munichdiff}. 
The other condensate is assumed to have a different 
interaction strength, and hence it may still be in
the superfluid phase with Poissonian number fluctuations on each site,
or it may have reduced fluctuations due to the on site repulsion. 
A detailed presentation of this process was 
given recently \cite{Hetero}.
When the optical lattice potential is raised the atomic wave functions
become more localized in the potential minima, and thus the
collisional interaction between atoms increases and the tunneling rate
between wells decreases. From the
time dependendent shape of the potential for the atoms one determines
the on-site interaction strengths 
$U_a,\ U_b$ and $U_{ab}$ between atoms of the 
same and different species $A$ and $B$
and the interwell tunneling couplings $J_a,\ J_b$,  but for simplicity
we shall just assume
the values shown in the upper panel of Fig.1. 
The ratio of approximately 0.6 between $U_a$ and $U_b$ 
corresponds to the ratio between the free space s-wave scattering
lengths of Rb and K atoms. The tunneling coupling of both species
are taken to follow the same lower curve.
These curves  are in qualitative agreement with the 
dependence of the interaction strength and tunnel coupling for
a single species evaluated  in \cite{Jaksch02}, where the  
energy and frequency unit $J_0$, on the order of a tenth of the atomic
recoil shift, is introduced ($\hbar=1$).
Using a symmetry breaking Gutzwiller ansatz, we assume that the
full many body system is in a product state over all sites, where each
site is described by the state vector 
$|\psi\rangle=\sum_{n_A,n_B} a_{n_A,n_B}|n_A,n_B\rangle$, 
with $n_i$ the number of atoms of species $i=A,B$.
We have solved the self-consistent equations for 
the amplitudes $a_{n_A,n_B}$, and we have 
verified that the two species undergo the Mott insulator transition 
at different times, see also \cite{Hetero}. The lower panel of 
Fig. 1. shows typical results of a numerical solution of the problem.
The solid lines
show the variance of the populations (the mean 
values are unchanged and equal to the initial variance
where we assume Poisson statistics). The dashed curves are obtained
by neglecting the cross coupling between the two species, and 
as in \cite{Hetero} we verify  that this coupling has 
little effect on the population statistics of the two species in the 
lattice wells. 

\begin{figure}[htbp]
\includegraphics[width = 80 mm]{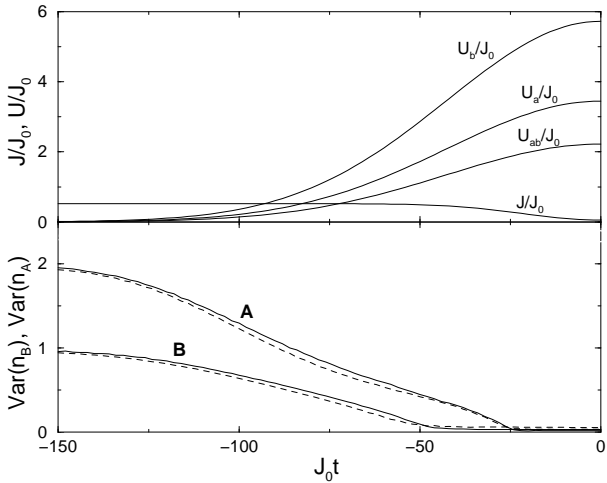}
\caption{Mott insulator transition for two atomic species.
The upper panel shows 
the time dependent coupling constants and 
the tunneling coupling, all in units of the constant $J_0$. 
The lower panel shows the variance of populations of the two atomic
species.  The dashed curves are obtained by neglect of the 
cross coupling term $U_{ab}$. 
.} 
 \end{figure}

Numerous studies have addressed the
prospects of forming molecules from ultra-cold atoms both theoretically
and experimentally \cite{Molecules}.
It has been
suggested to trap the atoms in lattices with low occupation number, 
and either make explicit use of the Mott insulator transition
\cite{Jaksch02,Hetero}, or simply work with low mean occupation of each
site \cite{Esslinger} to form the molecules in an optimally controlled 
manner.
We suggest to build on these proposals and to drive 
the  photoassociation process on each 
lattice site to form a heteronuclear molecule with a single atom
from each of the two species.  We are more interested in the process than
in the final molecular state, and hence a sufficiently long lived excited 
vibrational state will suffice as molecular state of the process.
The coherent photoassociation process is described by
the effective, second quantized, Hamitonian
\begin{equation}
H= \chi a_Aa_Ba^\dagger_M + \chi^*a_A^\dagger a_B^\dagger a_M
\label{ham}
\end{equation}
where $a_i$ is the annihilation operator of atoms and molecules $(i=A,B,M)$.
When restricted to the states with initially precisely one B atom,
we can rewrite this Hamiltonian as
\begin{equation}
H= \chi a_A\sigma^\dagger + \chi^*a_A^\dagger \sigma,
\label{hamjc}
\end{equation}
where the Pauli spin lowering operator $\sigma=a^\dagger_Ba_M$ represents the
two-state transition operator
$|1_B,\ 0_M\rangle\langle 0_B,\ 1_M|$, ($\sigma$ turns an AB molecule
into a B atom, i.e. , it does not conserve the number of A atoms,
and it appears only in combination with the creation operator
$a^\dagger_A$).  The Hamiltonian couples pairs of levels
\begin{equation}
|n_A,\ 1_B,\ 0_M\rangle \leftrightarrow 
|(n-1)_A,\ 0_B\ 1_M\rangle,
\label{pa}
\end{equation}
and it thus follows that there will be an oscillatory exchange of the 
single B atom between being part of the molecule and being an
independent atom, in complete analogy with the Jaynes-Cummings 
transfer of excitation between a two-level system and an oscillator
mode.

This oscillation
will undergo  collapses and revivals, because the coupling strength
betwen the two mentioned states is proportional to $\sqrt{n_A}$,
and hence the different number state components of A atoms
will give away an atom to the association process
at different frequencies. An experiment
that counts the total number of B atoms as a function of duration
of the photoassication process, can thus teach us about the number
statistics of the A atoms!

So far, the Mott insulator transition has been demonstrated
experimentally by the loss of interference between different
lattice sites, and by a gap in the excitation spectrum 
\cite{Munichdiff}. The Jaynes-Cummings dynamics presents a unique
application of a quantum optical method for  the study of this 
many-body problem.  We suggest to turn on slowly the
lattice potential, so that the B atoms undergo the transition to the Mott
phase at a time where the A atoms  are still in an unknown state. At
this time we hold the lattice potential and we turn on the 
photoassociation process, i.e., the dynamics
driven by the Jaynes-Cummings Hamiltonian (\ref{hamjc}).
The probability to find a B atom on any lattice site
is given by the square of the Rabi oscillation amplitudes, weighted by the 
$n_A$ probability distribution $p(n_A)$:
\begin{equation}
\langle n_B\rangle =\sum_{n_A} p(n_A) \cos^2(\chi\sqrt{n_A} t)
\label{collapse}
\end{equation}
This function is illustrated in Fig. 2. which shows the mean value of 
$n_B$ as a function of time in case of coupling to a Poissonian
distribution with on average two A atoms per lattice site (upper panel),
a sub-Poissonian distribution of A atoms achieved at $J_0t=-50$, cf. Fig 1, 
during application of the periodic potential (middle panel), and an 
$n_A=2$ (Mott insulator) number state (lower panel). 
For time sequences like the ones in Fig.2, one can use
Eq.(\ref{collapse}) to obtain a fit, which produces the complete
$p(n_A)$ distribution. 
By counting the total number of B atoms, without requiring the
experimental ability to detect atoms with high efficiency on a
microscopic spatial scale, we thus obtain the precise
number statistics at each site of the A atoms.
A similar procedure was initially applied for trapped ions \cite{Meekhof},
where the internal state of the ion was used to monitor the population
of the harmonic oscillator levels of motion. Many repeated experiments
on the same system  were necessary to probe the  state of the ion at
different durations $t$ of the dynamics and with good precision.
We also have to operate the process many times with different durations
of the photoassociation process, but since the experiment 'repeats
itself' in parallel in every lattice site, 
determination of the total number of B atoms in
the entire system suffices to give the mean value of $n_B$. 
To take into account both deviations from the 
Lamb-Dicke limit and decoherence, in Ref.\cite{Meekhof} the ideal 
Jaynes-Cummings dynamics (\ref{collapse}) was replaced by 
a more general formula
\begin{equation}
\langle n_B\rangle =\sum_{n_A} p(n_A) \cos^2(\Omega_{n_A} t)
e^{-\gamma_{n_A} t}
\label{collapseW}
\end{equation}
with fitting parameters $p(n_A)$, $\Omega_{n_A}$ and $\gamma_{n_A}$.
Similar extra handles on the theory may be useful in our matter-wave 
analog where, e.g.,  atom-atom interactions may cause an
$n_A$-dependence of the atomic wave function and hence of the
photo-association coupling rate, and where various
mechanisms, such as inhomogeneities over the size of the lattice,
may cause decoherence.

\begin{figure}[htbp]
\includegraphics[width = 80 mm]{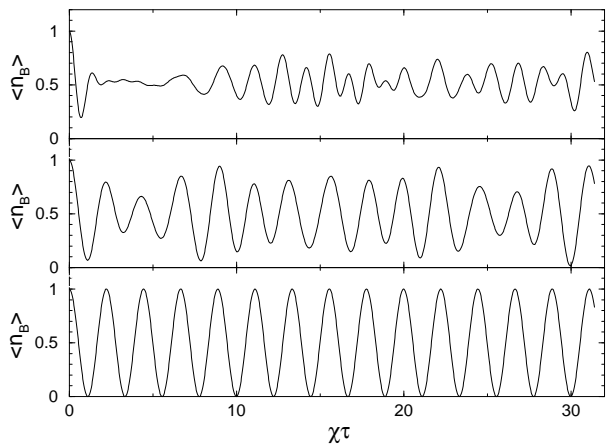}
\caption{Time dependence of the mean number of B atoms per
lattice site during photoassociation/-dissociation. The upper panel 
shows the results for a Poissonian
distribution with an average of two A atoms per site; the middle
panel shows the results for the distribution of A atoms at 
$J_0t=-50$ in Fig. 1, where the Mott insulator transition
has not yet taken fully place for the A atoms; and, the lower panel
shows the results for an $n_A=2$ number state.} 
\end{figure}

In case one can only effectively control the Mott-insulator transition
of one species (the A atoms), we note that for a sufficiently low 
density of B atoms, every lattice site will be  occupied by a single or
none of these atoms,
and hence the mean value of $n_B$ will ocillate in the same
way as in (\ref{collapse},\ref{collapseW}) but multiplied with the 
mean number of B atoms per site.  

As mentioned in the introduction of the Letter, the Jaynes-Cummings 
Hamiltonian has been applied extensively in quantum optics, and the
range of interesting phenomena studied is very wide. 
Let us turn to the possibilities for synthesis and studies of 
particular quantum
states of matter by means of the effective Jaynes-Cummings Hamiltonian.
Studies of the collapse and revival of the diffraction pattern of
lattice trapped atoms have already been made \cite{Munichrevival,
Yalediff},
and we propose to use the Jaynes-Cummings dynamics to produce a
Schr\"odinger cat like state of the A atoms, and to probe its coherence
properties.  The photoassociation process can be driven off resonance 
with a detuning $\delta$ in which case
the two-state transition (\ref{pa}) does not take place, but the coupling
(\ref{hamjc}) perturbs the energy of the state 
$|n_A, 1_B\rangle$ by the amount $\Delta E_{n_A}=
|\chi|^2n_A/\delta$. If the A atoms are in a coherent state
$|\alpha\rangle$, this state  will
undergo a simple phase rotation,$|\alpha\rangle \rightarrow 
|\alpha e^{i\phi}\rangle$ 
due to this coupling. Imagine now
that the B atom has a level structure, so that it can be initally
prepared in a superpostion of the active state $|B\rangle$  
which experiences the
coupling discussed above, and a passive state $|B'\rangle$ 
for which the photoassociative coupling vanishes completely. This 
implies that the Jaynes-Cummings Hamiltonian produces a material 
Schr\"odinger cat like state of the two species,
\begin{equation}
|\Psi\rangle = a|\alpha e^{i\phi}\rangle\otimes |B\rangle
+b|\alpha\rangle\otimes|B'\rangle,
\label{cat}
\end{equation}
very analogous  to the production of a similar state in cavity QED 
experiments \cite{ENS-cat}.
By driving the transition between the active and the passive level
in the B atom, we can subsequently monitor the interference of 
the two coherent components. The production of a mesoscopic superposition 
state can thus be verified experimentally.
In a gauge invariant formulation of the problem, there is of course not
a coherent state and mean atomic field in a single site, but a
total number state of 
the entire atomic ensemble may be well approximated by a coherent state,
and when this is split into a product of coherent states on each 
lattice site, a total number projection operator produces a very entangled
state of the atomic components on each site \cite{Esslinger},  but physical 
observables will be correctly handled by the coherent state ansatz
\cite{Klaus}.

The cavity QED implementation of the Jaynes-Cummings
Hamiltonian allows for classical excitation of both the
two-level system and  the oscillator, and it operates 
experimentally with the injection of a beam of 
excited two-level atoms, who can give off their excitation energy to
the oscillator degree of freedom. In the ion trap
implementation, both the oscillator and the two-level
system can be excited independently, and by changing the coupling
frequency an ``anti" Jaynes-Cummings Hamiltonian 
$\chi a \sigma + \chi^* a^\dagger \sigma^\dagger$, and a 
``two-phonon" Hamiltonian $\chi a^2 +\chi^* (a^\dagger)^2$ can be
implemented to produce various non-classical states \cite{Meekhof}. 
Tomographic reconstruction of the motion of a trapped
ion has been demonstrated in \cite{Leibfried}. That measurement scheme
made use of a classical driving to displace the oscillator by the complex
amplitude $\alpha$ and subsequently the populations of the various
oscillator states were probed by the two-level population oscillations
(\ref{collapse},\ref{collapseW}) as  described above. 
The position-momentum Wigner function was then obtained by an
analytical formula \cite{Leibfried} from the 
populations recorded with different displacement $\alpha$.

It will add an experimental level of difficulty to
couple atoms coherently from a reservoir condensate into the 
lattice potentials, but with the current progress in transfer and loading
of degenerate gasses into microfabricated structures \cite{Micro}, some
possibilities are certainly open. 
Without continuous loading of the lattice, there is another possibility
to mimic the quantum optical effects driven by external, classical  fields. 
We illustrate this with a protocol for full quantum state tomography 
of the matter wave field of the A atoms: 
Since we cannot accomplish coherent displacements of the atomic field
we suggest to use A atoms with two internal states, the active state 
$|A\rangle$ and a passive state $|A'\rangle$ which is not affected 
by the photoassication process. We can then use the $|A'\rangle$ population 
as a 'local oscillator', and as long as the A'-population is large, 
coherent transfer of atoms between the two atomic 
states is well described by a coherent displacement of the A atom oscillator.
In fact, this internal state
coupling is correctly described by means of a rotation of the Dicke 
collective spin of the atoms, and the detection of the population
statistics of the $|A\rangle$-state atoms by means of the 
Jaynes-Cummings dynamics can be used for perfect collective spin
tomography: If there is a total of  $N$ atoms populating states 
$|A\rangle$
and $|A'\rangle$ in every lattice site, it takes $4N+1$ effective 
spin rotations followed by population measurements  to reconstruct 
the full collective spin density
matrix, following the protocol applied to large hyperfine 
manifolds in Ref.\cite{Jessen}. 

In summary, we have suggested how to realize a purely atomic
Jaynes-Cummings Hamiltonian and how to use it both to 
generate quantum optical states of matter and to study various
many-body properties of the system.  To isolate the
different effects and mechanisms, we did not present studies of the 
joint effect of interactions, inter-well tunneling and the
photoassociation process. It is clear that these effects in
combination may provide wide possibilities 
for detection and production of quantum states.
It has been estimated in \cite{Jaksch02, Hetero} that one can reliably
transfer atoms coherently between the trapped ground states of the
system, but it is of course also an interesting possibility to use
the resonance conditions for the photoassociation lasers to 
excite and study collective excitations of the atoms.
We have focussed on lattice trapped atoms, where the 
coherent Jaynes-Cummings dynamics is most readily realized, and 
where the 
existence of many replica of the same quantum systems provides effective
means for detection of generic quantum optical behavior of the matter wave
fields. Within this framework, one can also consider multiple species
and multi-level systems in analogy with multi-mode and multi-level
generalizations of the Jaynes-Cummings Hamiltonian.
The dynamics is not restricted to lattices, and as an extension 
to the work we imagine that a collection of distinguishable, e.g.,
spatially localized B atoms immersed in a larger trapped condensates 
of A atoms can serve to 
induce a host of interesting effects in analogy with the more
general interaction of light and two-level atoms.  An appealing
possibility would, for example, be to look for  a
matter wave analogue of the Mollow fluorescence triplet, 
when an atomic laser beam interacts with an ensemble of B 
atoms and molecules can dissociate to atomic states with higher 
and lower energy than the energy of the incident beam.

Comments on the manuscript by Michael Budde are gratefully acknowledged.

\end{document}